\documentstyle[12pt]{article}

\begin{document}

\begin{titlepage}

\begin{center}
{\Large \bf  Long distance contribution to
$B^- \to K^- K^- \pi^+$, - a searching ground mode for new physics}\\
\vspace{1cm}
{\large \bf S. Fajfer$^{a,b}$,  P. Singer$^{c}$\\}

{\it a) J. Stefan Institute, Jamova 39, P. O. Box 3000, 1001 Ljubljana,
Slovenia}
\vspace{.5cm}

{\it b)
Department of Physics, University of Ljubljana, Jadranska 19, 1000 Ljubljana,
Slovenia}
\vspace{.5cm}

{\it c) Department of Physics, Technion - Israel Institute  of Technology,
Haifa 32000, Israel}

\end{center}

\vspace{0.25cm}

\centerline{\large \bf ABSTRACT}

\vspace{0.2cm}
The decay $B^- \to K^- K^- \pi^+$ has been sugested as a test for
minimal supersymmetric standard model and
for supersymmetric models with R-parity
violating couplings, in view of its extreme smallness in the
standard model.
We calculate two long distance contributions to this decay, that
associated with $DD$ and $D\pi$ intermediate
states and
that induced by
virtual $D$, $\pi$ mesons. The branching ratio due to these
contributions is $6 \times 10^{-12}$,
which is somewhat smaller than
the standard model short distance result,
leaving this decay free for the search of new physics.

\end{titlepage}

The standard model (SM) of strong and electroweak interactions is \break
presently in very good shape. The experimental data agree with SM
and this continuing success is somewhat paradoxically, a
principal factor in the intensive search for physics beyond
the standard model. This search is conducted nowadays in various
sectors of particles phenomena. Among these, rare b decays is
considered to provide good opportunities for discovering new
physics beyond SM \cite{AM1}.

Among the rare decays studied so far $b \to s \gamma$  plays
prominent role. The measured rate in two different experiments is
$Br(b\to s \gamma) = [3.15\pm 0.35 (stat) \pm $ $
0.32(syst) \pm 0.26(mod)] \times 10^{-4}$ \cite{CLEO1} and
$Br(b\to s \gamma) = [3.11\pm 0.80 (stat) \pm $ $
0.72(syst)] \times 10^{-4}$ \cite{ALEPH}, to be compared
with the latest theoretical calculations within the SM giving
$Br(B\to X_s \gamma) = (3.32\pm 0.30)\times 10^{-4}$ \cite{CHET}.
The agreement with SM is impresive and it is doubtful that
a deviation could be detected, even when the above figures are
improved. This, in view of the fact that long-distance
(LD) contributions are
also present; although more difficult to calculate with good
accuracy, the existing estimates concur that these are
approximately $(5 - 10)\%$ of the short-distance (SD)
amplitude \cite{GAD1}.

An alternative approach to the identification of virtual effects
from new particles in $b$ decays like $b\to s \gamma$ and
$b\to s l^+ l^-$ is the consideration of rare decays which are
of negligible strength
in SM. In such cases the mere 
appearence of the decays at a rate much larger than  it is
possible in SM would be a clear sign of new
physics.

Recently, the decays $b \to ss \bar d$, $b \to dd \bar s$
were proposed \cite{PAUL} as ideal prototypes
of the latter method.
As shown in  Ref. \cite{PAUL}, the $b \to ss \bar d$ is
mediated in the SM by box-diagram and its calculation
results in a branching ratio nearly of $10^{-11}$, the exact value
depending on the relative unknown phase between t, c contributions  
in the box.
The $b \to d d \bar s$ branching ratio is even smaller by a
factor of about $10^2$, due to the relative $|V_{td}/V_{ts}|$
factor in the amplitudes. The authors of ref. \cite{PAUL} have also
calculated the $b \to ss \bar d$ transition in various
"beyond the SM" models. It appears that for certain plausible
values of the parameters, this decay
may proceed with a branching ratio
of $10^{-8} - 10^{-7}$ in the minimal supersymmetric standard model
and in two Higgs doublet models.

Moreover, when one considers
supersymmetric models with R-parity violating couplings, it turns out that
the existing bounds on the involved couplings of the superpotential
do not provide at present any constraint on the $b \to s s \bar d$
mode \cite{PAUL}. It has been pointed out in Ref. \cite{PAUL} that
the  hadronic channels most suitable for the search of
the $b \to s s \bar d$ transition are the $\Delta S = 2$ decays
$B^- \to K^- K^- \pi^+$ or $\bar B^0 \to K^- K^- \pi^+ \pi^+$.
The appropriate exclusive channel for $b \to dd \bar s$
transition would be $B^- \to K^+ \pi^-  \pi^-$.
At present there is no
published experimental limits on these modes.

In the analysis of Ref. \cite{PAUL} only the short-distance contributions
were considered in detail. However, it is well known that long-distance
contributions which are associated with low-lying intermediate hadronic
states \cite{WOLF} are also present in particle transitions. As we
mentioned above, such contributions to $B \to X_s \gamma$ are rather small.
However, each specific decay mode requires the estimation of its
LD contribution;
this is imperative, since only when a  trustworthy estimate
of such contributions is available one may
proceed to compare the specific
transition to the theoretical SM treatment or use it for revealing new physics.
This necessity
is best exemplified by known occurences in K - physics \cite{PAUL1}:
in some decays like $K^+ \to \pi^+ \pi^0 \gamma$,
$K^+ \to \pi^+ l^+ l^-$ the SD contribution is obscured by LD contributions
while for $K_L^0 \to \pi^0 \nu \bar \nu$,
$K^+ \to \pi^+ \nu \bar \nu$, the LD contributions are considerably smaller
than the standard model short-distance amplitude \cite{LITT}.
The calculation of long-distance contributions to a specific process is not
based  on a well-defined theoretical procedure. Clearly, the intermediate
states are the main contributions to this part of the
amplitude.  However, the technique of their inclusion, as well
as the choice of relevant states will influence the final result.
We shall rely on the accumulated experience from the treatment of long-distance
contributions to various processes, like $K - \bar K$
transition \cite{WOLF}, $\bar D  -  D$ transition \cite{DONOG},
$K^+ \to \pi^+ \nu \bar \nu$ \cite{SEHGAL},
$B \to X_s \gamma$ \cite{GAD1} 
and $B_s \to \gamma \gamma$ \cite{ELLIS} decays, in order to formulate
our approach to the $B^- \to K^- K^- \pi^+$ process at hand.

We include two contributions in the calculation of the long distance
amplitude $B^- \to K^- K^- \pi^+$: (I) the box diagram, shown in Fig. 1,
which is essentially the LD analog of the
SD calculation in the standard model \cite{PAUL} of
the $b \to s s \bar d$ transition. (II)
 the contribution of virtual $"D^0"$
 and $"\pi^0"$ mesons, via the chain $B^- \to K^- "D^0"("\pi^0")$
 $\to K^- K^- \pi^+$. This contribution arises
 as a sequence of two $\Delta S = 1$ transitions and may lead to
 final $K^- K^- \pi^+$ state as well. It is therefore
 necessary to have an estimate of its relevance vis - \`a - vis
 the "direct" $\Delta S = 2$ transition. Let us consider firstly the amplitude
arising from
 (I). The two diagrams (a) and (b) express the Glashow -
 Iliopoulos - Maiani symmetry, so that the decay amplitude vanishes in
  the limit $m_c= m_u$ $(m_D = m_{\pi})$. Since each diagram contains two
  $W'$s, it is related to several semileptonic processes with
  one virtual meson.
  Thus diagram (a) relates to $D^0 \to K^- e^+ \nu_e$,
  $B^- \to D^0 e^- \bar \nu_e$ and
  $D^0 \to \pi^+ e^- \bar \nu_e$  involving the product
  $V_{cb} V_{cs}^* V_{cd} V_{cs}^*$ and diagram (b) relates
  $D^0 \to K^- e^+ \nu_e$,
  $B^- \to D^0 e^- \bar \nu_e$,
  $K^- \to \pi^0 e^- \bar \nu_e$ and $\pi^+ \to \pi^0 e^+ \nu_e$
involving the product  $V_{cb} V_{cs}^* V_{ud} V_{us}^*$.

The transition probabilty   is given by
\begin{eqnarray}
< K^- K^- \pi^+ | {\cal S }| B^- >_{box} =  (\frac{ig}{2\sqrt{2}})^4
V_{cb} V_{cs}^* V_{cd} V_{cs}^* \int d^4 q_1 d^4q_2 d^4Q_2 d^4Q_1&&\nonumber\\
\{ \delta^4 ( Q_1 + q_2 - k_2)
\delta^4 (p_B -q_1 - Q_1)  \delta^4 (Q_2 - p_\pi - q_2)
\delta^4 (q_1 - Q_2 - k_1) &&
\nonumber\\
 < D^0 | (\bar c b)^{\nu}  |B^-> \frac{i}{q_1^2 -m_D^2}
< K^- | (\bar sc)^{\alpha}  |D^0>
\frac{ -i (g_{\nu \alpha} -
Q_{1\nu} Q_{1\alpha} /M_W^2)}{Q^2_1 - M_W^2}& &\nonumber\\
\frac{ -i(g_{\mu \beta} -
Q_{2\mu} Q_{2\beta} /M_W^2)}{Q^2_2 - M_W^2}[
< K^- | (\bar sc)^{\mu}  |D^0>
\frac{i}{q_2^2 -m_D^2}
< \pi^+ D^0  |(\bar c d)^{\beta} |0>&& \nonumber\\
 - < K^- | (\bar su)^{\mu}  |\pi^0>
\frac{i}{q_2^2 -m_\pi^2}
< \pi^+ \pi^0  |(\bar u d)^{\beta} |0>]&&\nonumber\\
+  \delta^4 ( Q_1 - q_2 - k_2)
\delta^4 (p_B -q_1 - Q_1)  \delta^4 (Q_2 - p_\pi + q_2)
\delta^4 (q_1 - Q_2 - k_1) &&
\nonumber\\
 < D^0 | (\bar c b)^{\nu}  |B^-> \frac{i}{q_1^2 -m_D^2}
< K^- | (\bar sc)^{\mu}  |D^0>
\frac{ -i (g_{\nu \alpha} -
Q_{1\nu} Q_{1\alpha} /M_W^2)}{Q^2_1 - M_W^2}& &\nonumber\\
\frac{ -i(g_{\mu \beta} -
Q_{2\mu} Q_{2\beta} /M_W^2)}{Q^2_2 - M_W^2}[
< K^- \bar D^0 | (\bar sc)^{\alpha}  | 0>
\frac{i}{q_2^2 -m_D^2}
< \pi^+  |(\bar c d)^{\beta} | \bar D^0 >&& \nonumber\\
 - < K^- \pi^0| (\bar su)^{\alpha}  |0>
\frac{i}{q_2^2 -m_\pi^2}< \pi^+ |(\bar u d)^{\beta}| \pi^0 >]
 +   (k_1 \leftrightarrow k_2)\},&&
 \label{eq1}
\end{eqnarray}
where $(\bar q_j q_i)^{\alpha}$ stands for  $\bar q_j
\gamma^{\alpha} (1- \gamma_5) q_i$, while the rest of
the notation is defined in
Figure 1. The first part comes out from the diagrams on Figure 1,
while the second results from the crossed diagrams.
The calculation of (\ref{eq1})
depends on the
matrix elements 
$< D^0 | (\bar c b)^{\nu}  |B^->$,
$< K^- | (\bar sc)^{\mu}  |D^0>$,
$< \pi^+ |(\bar c d)^{\beta} |\bar D^0 >$,
$< K^- | (\bar su)^{\mu}  |\pi^0>$ and
$< \pi^0  |(\bar u d)^{\beta} |\pi^->$.
Since only pseudoscalar states appear, we have to deal
with transitions between such states induced by the
vector current only,
\begin{eqnarray}
< P'(p') | \bar q_j \gamma^{\mu} q_i |P(p)> =
f_+(q^2) (p^{\mu} + p'^{\mu})
+  f_-(q^2) (p^{\mu} - p'^{\mu}),&& 
\label{eq2}
\end{eqnarray}
which may be rewritten as \cite{WSB1}
\begin{eqnarray}
< P'(p') | \bar q_j\gamma^{\mu} q_i|P(p)> & = &
F_1(q^2) (p^{\mu} + p'^{\mu} - \frac{m_P^2 - m_{P'}^2}{q^2}
(p^{\mu} - p'^{\mu})) \nonumber\\
&+ & F_0(q^2) \frac{m_P^2 -m_{P'}^2}{q^2}
(p^{\mu} - p'^{\mu}), 
\label{eq3}
\end{eqnarray}
where $F_1$ and $F_0$ contain the contribution of vector and scalar
states respectively and $q^2 = (p - p')^2$. Also, $F_1(0) = F_0(0)$
\cite{WSB1}. For these form factors, one usually assumes pole
dominance \cite{SEHGAL,WSB1,JURE,WSB2}
\begin{eqnarray}
F_1(q^2) & = & \frac{F_1(0)}{ 1 - \frac{q^2}{m_V^2}}; \enspace
F_0(q^2)  =  \frac{F_0(0)}{ 1 - \frac{q^2}{m_S^2}}
\label{eq4}
\end{eqnarray}
and in order to simplify, we shall take $m_V = m_S$, from which results
$f_- (q^2) = 0$.
We shall assume that one can safely take $f_+(q^2)\simeq 1$
\cite{DONOG} and the limit $Q^2_1$ , $Q_2^2 \ll M_W^2$, which then
leads to a more tractable expression for the real part
of the amplitude
\begin{eqnarray}
{\cal A}_{r}^{box}(B^-(p_B) \to K^-(k_1) K^-(k_2) \pi^+(p_\pi))   =
\frac{G^2}{16\pi^4 }
V_{cb} V_{cs}^* V_{cd} V_{cs}^* \int d^4 q_1
&&\nonumber\\
\{\frac{1}{q_1^2 -m_D^2} \frac{1}{(q_1 -k_1 -p_\pi)^2 - m_D^2}
[ ( - m_B^2 + 2 k_2 \cdot p_B) (m_K^2 + 2 k_1 \cdot p_\pi) &&\nonumber\\
  + q^2_1 (m_K^2 + 2 k_1 \cdot p_\pi + m_B^2 - 2 k_2 \cdot p_B)
+ 2 k_2 \cdot q_1 (m_K^2 + 2 k_1 \cdot p_\pi)&& \nonumber\\
 + 2 p_{\pi} \cdot q_1 2 k_2 \cdot q_1 - 2 k_2 \cdot q_1 q_1^2 +
 2 p_\pi \cdot q_1 q_1^2
+ 2  p_\pi \cdot q_1 ( - m_B^2  + 2 p_{\pi} \cdot p_B)   - q_1^4 ] & &
 \nonumber\\
-\frac{1}{q_1^2 -m_D^2} \frac{1}{(q_1 -k_1 - p _\pi)^2 - m_\pi^2}
[ (- m_B^2 + 2 k_2 \cdot p_B) (m_K^2 + 2 k_1 \cdot p_\pi) &&\nonumber\\
  + q^2_1 (m_K^2 + 2 k_1 \cdot p_\pi + m_B^2 - 2 k_2 \cdot p_B)
+ 2 k_2 \cdot q_1 (m_K^2 + 2 k_1 \cdot p_\pi)&& \nonumber\\
 + 2 p_{\pi} \cdot q_1 2 k_2 \cdot q_1 - 2 k_2 \cdot q_1 q_1^2 +
 2 p_\pi \cdot q_1 q_1^2
+ 2  p_\pi \cdot q_1 ( - m_B^2  + 2 p_{\pi} \cdot p_B)  - q_1^4 ] & &
 \nonumber\\
 + (k_1 \leftrightarrow k_2) \}, 
\label{eq5}
\end{eqnarray}
where $G = {\sqrt 2} g^2/(8 M_W^2)$.
The separate contributions of $DD$ and $D\pi$
intermadiate states diverge as fourth power. However,
the GIM  cancellation acts in such a way as to decrease the degree of
divergence and finally the integral in (\ref{eq5}) will give a
quadratic divergence. Similar situations were encountered in previous
LD calculations
\cite{WOLF,SEHGAL}. We note that the explicit inclusion of the
pole-type form factors
for $f_+ (q^2)$ would reduce the degree of divergence, in such a case,
however, the evaluation of the integrals becomes very cumbersome and
this effort is not justifed since as it will
turn out the contribution of the real
part is essentially negligible in comparison to
that provided by the imaginary part. The integrals in (\ref{eq5})
are calculated by using Feynman parametrization. The final result
for the decay
rate is
\begin{eqnarray}
\Gamma (B^- \to K^- K^- \pi^+) & = &  \frac{1}{2 (2 \pi)^3 32 m_B^3}
\int_{(m_\pi +m_K)^2}^{(m_B -m_K)^2} ds_2 \int_{(s_1)_1}^{(s_1)_2}
ds_1 |{\cal A}|^2,
\label{eq6}
\end{eqnarray}
where
\begin{eqnarray}
(s_1)_{1,2} & = & m_K^2 + m_{\pi}^2 - \frac{1}{2 s_2} [
(s_2 - m_B^2 + m_K^2) (s_2 + m_{\pi}^2 -m_K^2) \nonumber\\
 &\pm &
 \lambda^{1/2} (s_2, m_B^2,m_K^2)
\lambda^{1/2}(s_2, m_\pi^2,m_K^2)]\nonumber\\
\end{eqnarray}
 and
$\lambda (a,b,c) = a^2 + b^2 + c^2 - 2 ab - 2 a c - 2 a b$.
The ${\cal A}_{r}^{box}$ denotes the leading term of the amplitude,
which results after using
the primitive cut-off regularization:
\begin{eqnarray}
{\cal A}_{r}^{box}  (B^- \to K^- K^- \pi^+) &\simeq& G^2
V_{cb} V_{cs}^* V_{cd} V_{cs}^*\frac{1}{16 \pi^2}
\Lambda^2 (m_D^2 - m_\pi^2).
\label{eq7}
\end{eqnarray}
There is obviously the uncertainty in the value
to be taken for $\Lambda$. The momentum in the box cannot
exceed $m_B$,
and by taking $\Lambda \simeq 10$ $ \rm GeV$ we obtain
\begin{equation}
BR(B^- \to K^- K^- \pi^+)^{(box)}_{(r)} \simeq 8 \times 10^{-15}
\label{eq8}
\end{equation}
for the real part of this contribution, using
$\Gamma (B^- \to all) = 4 \times 10^{-13}$ $ \rm  GeV$
\cite{CASO}.

Turning now to the imaginary part of the
$B^- \to K^- K^- \pi^+$
amplitude provided by the $DD$ and $D\pi$
intermediate  states, it is given by
\begin{eqnarray}
{\cal A}_{i}^{box}(B^- \to K^- K^- \pi^+)   =
- \frac{G^2}{32 \pi^2}
V_{cb} V_{cs}^* V_{cd} V_{cs}^*  && \nonumber\\
 \int {d^4 q_1} \delta(q_1^2 - m_D^2) [ \delta((q_1- k_1 - k_2)^2 - m_D^2) -
\delta((q_1- k_1 - p_\pi)^2 - m_{\pi}^2) ] &&\nonumber\\
\{- q_1^4 + - 2 k_2 \cdot q_1 q_1^2  + 2 p_\pi \cdot q_1 q_1^2
+q_1^2 (m_K^2 + 2 k_1 \cdot p_\pi +  m_B^2 - 2 k_2 \cdot p_B)
&& \nonumber\\
+ 2 p_\pi \cdot q_1 2 k_2 \cdot q_1  +2 p_\pi \cdot q_1
(- m_B^2 + 2 k_2 \cdot p_B)&& \nonumber\\
+ 2 k_2 \cdot q_1 (m_K^2  +2 p_\pi \cdot k_1) +
(- m_B^2 + 2 k_2 \cdot p_B)(m_K^2 + 2 k_1 \cdot p_\pi) &&\nonumber\\
 + (k_1\leftrightarrow k_2)\}.&&
\label{eq9}
\end{eqnarray}

Introducing now $s_1= (p_B - k_1)^2 = $ $( k_2 + p_\pi)^2$
and $s_2= (p_B - k_2)^2 = $ $( k_1 + p_\pi)^2$ one arrives at
\begin{eqnarray}
{\cal A}_{i}^{box}(B^- \to K^- K^- \pi^+)   =
- \frac{G^2}{32 \pi^2}
V_{cb} V_{cs}^*V_{cd} V_{cs}^* && \nonumber\\
\times \lbrace F(m_D^2,s_1,s_2) - F(m_{\pi}^2,s_1,s_2)
+ (s_1\leftrightarrow s_2) \rbrace,&&
\label{eq11}
\end{eqnarray}
with
\begin{eqnarray}
F(m_P,s_1,s_2) =
\frac{\lambda^{1/2} (m_D^2,m_P^2,s_1)}{m_D^2- m_P^2 +s_1}&&
 \nonumber\\
\lbrace - m_D^4
+  m_D^2 (2s_1 - \frac{3}{2}m_K^2  - \frac{3}{2}
m_{\pi}^2 - \frac{1}{2} s_2) +  (s_1 - m_{\pi})
(-s_1 + m_K^2)
 &&\nonumber\\
 + (s_1\leftrightarrow s_2)\rbrace. &&
 \label{eq12}
\end{eqnarray}
Using the expression (\ref{eq6}) for the decay width we find
\begin{equation}
BR(B^- \to K^- K^- \pi^+)_{(i)}^{(box)} \simeq 6 \times 10^{-12}.
\label{eq13}
\end{equation}

We proceed now to estimate the second  possibility for a LD
part which may lead to a final $K^- K^- \pi^+$ state. This possibility is
expressed
as two consecutive two-body nonleptonic transitions (see Fig. 2)
in which the connecting single
particles $D^0$ and $\pi^0$  is virtual. An estimate of this
contribution requires the knowledge of the
$< B^-|{\cal H}_w|K^- "D^{0}">$, $<"D^{0}"|{\cal H}_w| K^- \pi^+>$
amplitudes for virtual $D^0$, which is lacking. For a virtual
$\pi^0$ existing estimates for $< "\pi^0"|{\cal H}_w| K^- \pi^+>$
indicate \cite{SEHGAL}  that it is smaller than the
physical amplitude in a certain region. We rely in our
estimation on the "physical" amplitudes $B^- \to K^- D$,
$D^0 \to K^- \pi^+$ \cite{WSB2}, keeping
in mind that this induces an amount of uncertainty. However,
the final numerical results will show that this is of no
consequence in the present problem.
In the diagram (2b) the $D^0$ may also be on the mass shell.
Therefore, we must exclude in our calculations the region
around physical $D^0$, which represents two $\Delta S = 1$
physical decays, $B^- \to D^0 K^-$ followed by
$D^0 \to K^- \pi^+$, since we are pursuing the
$B^- \to K^- K^- \pi^+ $ outside the resonance region. We shall
return to this point below.

The calculation of the virtual $D^0$ mediated part
of the amplitude
requires the use of the effective nonleptonic Lagrangian.
The part relevant for the present calculation is
\begin{eqnarray}
{\cal L}_{LD} & =& - \frac{G}{\sqrt 2} \lbrace V_{cb} V_{us}^*
\lbrack a_1^{(b)}
(\bar c b)^{\mu} (\bar s u)_{\mu}
+ a_2^{(b)}(\bar c u)^{\mu} (\bar s b)_{\mu}\rbrack \nonumber\\
& + & V_{cs} V_{ud}^* \lbrack a_1^{(c)}
(\bar c s)^{\mu} (\bar d u)_{\mu}
+ a_2^{(c)}(\bar c u)^{\mu} (\bar d s)_{\mu} \rbrack + h.c.\rbrace,
\label{eq14}
\end{eqnarray}
$V_{q_1 q_2}$ are CKM matrix elements and $a_1^{(c)}$, $a_2^{(c)}$,
 $a_1^{(b)}$ and  $a_2^{(b)}$ are effective Wilson coefficients (see
 Bauer et al., Ref. \cite{WSB2}) at the charm and beauty scales.
 We use factorization approximation \cite{WSB2} for the two
 parts of the $B^- \to K^- K^- \pi^+$ amplitude and the expression
 obtained in \cite{JURE} for the $< P_1 P_2| D>$ transition,
 \begin{eqnarray}
 M_{< P_1 P_2| D>} &= & \frac{G}{\sqrt{2}} C_{P_1P_2} i f_{P_2}
 F_0^{D\to P_1} (m_{P_2}^2) (m_D^2 - m_{P_1}^2).
 \label{eq15}
 \end{eqnarray}
 In (\ref{eq15}) $ C_{P_1P_2} $ contains CKM matrix elements and a
 Wilson coefficient. By using the explicit form of (\ref{eq15})
 we have neglected the small contribution from
 the annihilation part of the amplitude, which is proportional to
 $a_2^{(b)}$, $a_2^{(c)}$ \cite{JURE}.
 The part of the decay amplitude
 due to the $"D^0"$ pole is then given by
 \begin{eqnarray}
{\cal A}_{D^0}^{pole}(B^- \to K^- K^- \pi^+)   =
- \frac{G^2}{2}
V_{cb}V_{us}^*  V_{cs} V_{ud}^*a_1^{(b)} a_1^{(c)}f_K f_{\pi} && \nonumber\\
 \times F_0^{B D}(m_K^2) F_0^{DK} (m_{\pi}^2)
 \frac{  (m_B^2 - q^2) (q^2- m_K^2)}{q^2 - m_{D}^2 +
 i m_D \Gamma_D}.
 \label{eq16}
 \end{eqnarray}
 The decay width due to this contribution
 is given by
 \begin{eqnarray}
\Gamma (B^- \to K^- K^- \pi^+)  =  \frac{1}{2 (2 \pi)^3 32 m_B^3}
|C|^2 |F_0^{B D}(m_K^2) F_0^{DK}(m_{\pi}^2) |^2& & \nonumber\\
 \times \int_{(m_\pi +m_K)^2}^{(m_B -m_K)^2} ds
 \frac{  (m_B^2 - s)^2
 (s- m_K^2)^2}{(s - m_{D}^2)^2 + ( m_D \Gamma_D)^2}
\nonumber\\& & \nonumber\\
 \frac{1}{s} \lambda^{1/2} (s,m_B^2,m_K^2)
 \lambda^{1/2} (s,m_\pi^2,m_K^2), 
 \label{eq17}
\end{eqnarray}
with  $C = ({G^2}/{2})
 V_{cb}V_{us}^*  V_{cs} V_{ud}^* a_1^{(b)} a_1^{(c)}f_K f_{\pi}$,
for the resonance in a $s$ channel and the same
for the resonance in a crossed channel.

Using for $a_1^{(b)}$, $ a_1^{(c)}$, $ F_0^{DK}$ and
$F_0^{B D}$ the values of Bauer, Stech and Wirbel \cite{WSB2}
we calculate the virtual
$"D^0"$ contribution by deleting a width of $2 \Delta$ around
$D$ mass in the $s$ variable. The size of
$\Delta$ is related to the experimental accuracy of the $D$ -
determination in
the final $K^- \pi^+$ state. In the various experiments it
ranges between 1 and 10 $\rm MeV$. One should keep in mind that
average accuracy of $D^0$ - mass determination is $0.5$ $\rm MeV$
\cite{CASO}. Thus, in order to delete the physical $D^0$'s one must
take at least $\Delta =1$ $\rm MeV$. However, we shall check the
$\Delta$ dependence for a range of values to make sure that
our conclusions are not affected.

For $\Delta = 20$, $5$, $1$, $0.1$
$\rm MeV$ we find  the nonresonant  $D^0$ contribution to be
 \begin{eqnarray}
 BR(B^- \to K^- K^- \pi^+)^{pole}_{(D^0)}& = &(0.31; \enspace
 1.2; \enspace 6.2; \enspace 61) \times 10^{-15}.
 \label{eq18}
 \end{eqnarray}
 In all cases the result is much smaller than (13), though
 one should remember that $\Delta = (1-5)$ $\rm MeV$ is the realistic
 option.
 A similar calculation for $\pi^0$ intermediate contribution,
 i.e. $B^- \to K^- "\pi^0"$ $ "\pi^0"\to K^-\pi^+$ yields
 a value smaller by four orders of magnitude, especially as a
 result of CKM angles. The pole contribution is therefore
 considerablly smaller than the LD box contribution calculated
 with $DD$ and $D\pi$ intermediate states; thus the total
 branching ratio from all diagrams we included is
 \begin{eqnarray}
 BR(B^- \to K^- K^- \pi^+)_{LD}& = & 6 \times 10^{-12}.
 \label{eq19}
 \end{eqnarray}
As a check, we used our ${\cal A}_{(D^0)}^{pole}$ amplitude
to calculate $B^- \to K^- K^- \pi^+$ as given by decay via a
physical $D^0$ and we find a branching ratio of $7 \times 10^{-6}$.
This agrees very well with the experimental expectation of
$(9.9 \pm 2.8 )\times 10^{-6}$, obtained by using
$BR(B^- \to D^0 K^-)$
$ = ( 2.57 \pm 0.65 \pm 0.32)\times 10^{-4}$ \cite{CLEO2} and
$BR(D^0 \to K^- \pi^+) = 3.85  \times
10^{-2}$ \cite{CASO}.

A few remarks concerning our approximations. As we mentioned,
we have neglected the form factor dependence
in the calculation of ${\cal A}_{r}^{box}$. Their inclusion would
have decreased the degree of divergence. However,
in view of the smallness of the real part of the amplitude,
this neglect is of no consequence.
A possibly more serious uncertainty is caused by the
fact that we used only $DD$ and $D \pi$
intermediate states in the box
calculations. Additional intermediate states, all within the
physical region could be $D D^*$, $D^* D^*$, $D \rho$, $D^* \rho$,
$D \eta (\eta')$.
We did not consider these states for two main reasons: first,
there is no knowledge of the matrix elements and the required
form factors involved.
Moreover, the inclusion of
strongly decaying resonances $D^*$, $\rho$ as intermediate states is
questionable
and some of their effects are taken into
account by the form factors considered (\ref{eq4}).  We decided therefore to
ignore these contributions,
though we are aware of the possibilty that additional intermediate
states in Fig. 1 might
increase our result by a factor of, say, 2-3.
Finally, it is interesting to note that
our result, which indicates that the LD contribution
in the $B^- \to K^- K^- \pi^+$ decay is smaller or at most
comparable to the SM short-distance
contribution, fits into the general picture
of B decays.
This is in contrast to the situation in the strange and charm
sectors:
in K transitions the two contributions are
comparable in some cases, SD dominates in a few decays and LD in
many others; in $D$ - transitions LD contributions are generally
larger than the SD ones, except for the unusual case of
$B_c \to B_u^* \gamma$ decay \cite{FPS}. \\

To summarize, we have shown that the long - distance
contributions to $B^- \to K^- K^- \pi^+$ are smaller in the SM
than the short - distance box diagram,
and have the branching ratio in  a $10^{-12} - 10^{-11}$ range.
 This is a most welcome
feature since it strengthens the suitability of the
$B^- \to K^- K^- \pi^+$ decay as an ideal testing ground
for physics beyond the standard model, as originally
suggested in ref. \cite{PAUL}. We expect that this avenue will
be explored experimentally in the near future and we note that the first
analysis of this mode has just been completed by the OPAL
Collaboration \cite{OPAL} and an upper limit
of $1.29 \times 10^{-4}$ at $90\%$ confidence
level has been set for the branching ratio
of this decay. \\

This work has been supported in part by the Ministry of
Science of the Republic of Slovenia (SF) and by the Fund for Promotion
of Research at the Technion (PS). We acknowledge with thanks discussion
with Drs. Yoram Rozen and Shlomit Tarem on the experimental aspects of
the problem.
One of us (SF) thanks A. Ram\v sak and D. Veberi\v c for
their help in numerical calculations. \\

\newpage

{\bf Figure Captions}\\

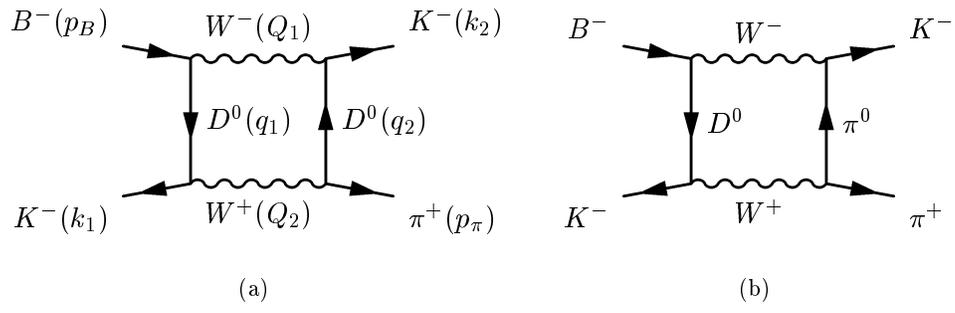
Fig. 1. Long distance box-diagram contributions to
$B^- \to K^- K^- \pi^+$. \\

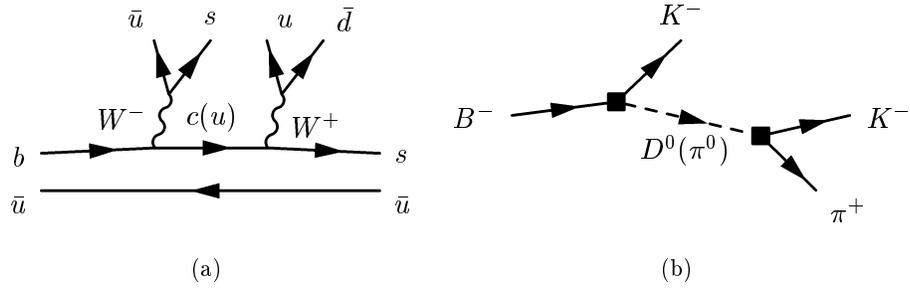
Fig. 2. Pole contributions to the long distance amplitude of
$B^- $ $\to K^- K^- \pi^+$, (a) quark picture, (b) hadronic picture.\\

\end{document}